\documentclass[a4paper,12pt]{article}
\pdfoutput=1
\usepackage{amssymb}
\usepackage{amsmath}
\usepackage{epsfig}
\usepackage{graphicx}
\usepackage{color,ulem}
\usepackage{hyperref}
\usepackage{kotex}
\usepackage{caption}
\usepackage{multirow}
\usepackage{comment}
\makeatletter

\@addtoreset{equation}{section}
\makeatother
\newcommand{\lambdabar}{{\mkern0.75mu\mathchar '26\mkern -9.75mu\lambda}}

\usepackage{hyperref}
\hypersetup{colorlinks,%
  citecolor=blue,%
  linkcolor=blue,%
  urlcolor=blue,%
}

\begin{document}
\begin{titlepage}
\vskip1cm
\begin{flushright}
\end{flushright}
\vskip0.25cm
\centerline{
\bf \large
Head-on Collisions of Fuzzy/Cold Dark Matter Subhalos
}
\vskip1cm \centerline{ 
  Hyeonmo Koo} 
\vspace{1cm}
\centerline{\sl  Physics Department,
University of Seoul, Seoul 02504 \rm KOREA}

 \vskip0.4cm
 \centerline{
\tt{(mike1919@uos.ac.kr)}}
  \vspace{2cm}
\centerline{ABSTRACT} \vspace{0.5cm}
{
\noindent
We perform head-on collision simulations of compact dark matter subhalos using distinct numerical methods for fuzzy dark matter (FDM) and cold dark matter (CDM) models.
For FDM, we solve the Schr\"odinger-Poisson equations with a pseudospectral solver, while for CDM, we utilize a smoothed particle hydrodynamics $N$-body code. 
Our results show that velocity decrease of subhalos is significantly greater in FDM model than in CDM, particularly at lower initial velocities, attributed to gravitational cooling-a unique mechanism of stabilizing in FDM with dissipating kinetic energy.
This stark contrast in energy dissipation between two DM models suggests that FDM may offer valuable insights into understanding the dynamic behaviors of DM during galaxy cluster collisions, such as those observed in the Bullet cluster and Abell 520.
These findings strongly suggest that FDM is not only capable of explaining these complex astrophysical phenomena but also serves as a compelling alternative to the traditional CDM model, offering resolutions to longstanding discrepancies in DM behavior.

\vspace{1.00cm}
\centerline{KEYWORDS} \vspace{0.5cm}
{
\noindent 
Fuzzy dark matter, Gravitational Cooling, Subhalo, Head-on collision
}
}

\end{titlepage}
\section{Introduction}\label{sec1}

Numerical simulations based on the cold dark matter (CDM) model, which is assumed to be collisionless, have been successfully explained the large-scale structure of the universe.
However, this model encounters significant challenges in explaining specific galactic and sub-galactic phenomena.
For instance, the expected sharp central density profiles in galactic halos do not match observational data, which show flatness in small galaxies \cite{Navarro:1995iw,deblok-2002,crisis}, known as the cusp-core problem.

Notably, two recent observations of galaxy cluster collisions, the Bullet cluster (1E0657-56) \cite{bullet} and Abell 520 \cite{abell520}, highlight discrepancies for the standard CDM framework.
Galaxy clusters are mainly composed of three distinct components: stars, baryonic gas, and DM \cite{Markevitch:2007dn}.
Since the stars in galaxies are sparse, they can be treated as collisionless, and are thus expected to move together with the DM after a cluster collision, while baryonic gases lag behind due to their electromagnetic interactions.
The observation of Bullet cluster seems to be consistent with this expectation \cite{bullet}.
However, in the Abell 520, gases and DM are centered in the core of the cluster, where stars have moved away \cite{abell520}.
This not only suggests that DM may exhibit collisional properties similar to gases, but also highlights the limitations of CDM model in accurately explaining the nuanced behaviors observed in galactic and sub-galactic structures.
Consequently, these problems underscore the need for alternative DM models that can more accurately account for such astrophysical phenomena across various scales.

An ultra-light bosonic scalar field with a particle mass $m \sim \mathcal{O}(10^{-22} {\rm \, eV})$ has recently emerged as an alternative to CDM \cite{Fuzzy,Hui:2016ltb,Schive:2014dra,Hui:2021tkt,Lee:2017qve}. 
This model has a typical de Broglie wavelength $\lambda_{\rm dB}\sim \mathcal{O}(1 \rm kpc)$ and thus exhibits wave-like behavior at galactic scales.
This behavior has led to the model being referred to as fuzzy dark matter (FDM), which could resolve the discrepancies between CDM predictions and observational data at these scales.
The formation of minimum-energy soliton configurations at the centers of galactic halos, characterized by flat core-like densities \cite{P.Mocz:2017}, provides insights into resolving the cusp-core problem.
Moreover, the FDM model is noted to offer potential solutions for several astrophysical phenomena--contradictory behaviors of DM observed in galaxy cluster collisions \cite{lee2008bec}, power-law relation between the mass of supermassive black holes in galaxies and the velocity dispersions of their bulges: called M-sigma relation \cite{lee:2015yws}.

A remarkable phenomenon in the FDM model is gravitational cooling, a relaxation process that ejects high-momentum modes of DM particles from a potential well, thereby carrying excessive kinetic energy away \cite{Guzman:2006yc, Bak:2018pda}. 
This mechanism is expected to play a crucial role in reducing velocities after interactions between FDM subhalos, differing from the effects of dynamical friction \cite{Lancaster:2019mde}.
Meanwhile, the CDM model relies solely on dynamical friction, as described by Chandrasekhar \cite{Chandrasekhar:1943ys}, which is based on the classical Newtonian dynamics.
Due to these distinct mechanisms, the dynamics of FDM subhalos are expected to significantly differ from those of CDM subhalos.

This study investigates the distinct dynamics of FDM and CDM subhalos during head-on collisions, with a focus on resolving discrepancies observed in phenomena like Abell 520.
Previous studies suggest that FDM, through mechanisms like gravitational cooling, may better explain the complex behavior of dark matter in such high-speed collisions. 
Such studies have numerically explored head-on collisions of FDM halos by incorporating luminous matter into the simulations \cite{Guzman:2016peo}, analyzing mergers at sufficiently low velocities \cite{Avilez:2018uot}, or extending the framework to multistate configurations \cite{Guzman:2018evm}.
Our work aims to strengthen this argument by providing further evidence that FDM could serve as a viable alternative to CDM.
Distinct numerical codes are used to simulate head-on collisions of subhalos for each DM model.
This study will illustrate how gravitational cooling in FDM leads to observably distinct outcomes compared to the dynamical friction effects in CDM.

In Section \ref{sec2}, we introduce the simulation setups of FDM and CDM system using different numerical codes.
We also review the basics of FDM physics required to set initial configurations. 
In Section \ref{sec3}, we present results of our simulations, such as snapshots and the change of the relative velocity of subhalos. 
In Section \ref{sec4}, we analyze the result of our simulations. This is based on the dynamical friction in both dark matter models and gravitational cooling in FDM.


\section{Simulation Setup of Head-on Collisions}\label{sec2}
In this section, we perform a series of head-on collision simulations involving two equal-mass DM subhalos within a cubic box of size $L=68\,\rm kpc$. 
Our primary focus is to analyze the differences in velocity changes after the collision between FDM and CDM subhalos.

We configure the initial conditions for subhalos in both DM models as follows. 
First, we set their mass to $M_s=2\pi\times 10^8\,M_{\odot}$, representing a typical mass scale for observed dwarf galaxies. 
Second, we fix their positions to $\mathbf{x}_{L}=\left(-\frac{L}{4},0,0\right)$ and $\mathbf{x}_{R}=\left(\frac{L}{4},0,0\right)$, placing their center-of-mass (COM) at the origin. 
Third, we set their velocities in the $x$-direction using 221 different initial relative speeds $v_i$, which has a range of $101.5 \ {\rm km/s} \lesssim v_i \lesssim 349.6 \ {\rm km/s}$, such that $\mathbf{v}_L=\left(\frac{1}{2}v_i,0,0\right)$ and $\mathbf{v}_R=\left(-\frac{1}{2}v_i,0,0\right)$.
All specific values, including box size, subhalo mass, initial positions, and velocities, are scaled to multiples of code units (e.g., $L=10\ell_c$, $M_s = 50\mathcal{M}$, and $v_i = 36v_c,\ ...,\ 124v_c$) within the Schr\"odinger-Poisson system, whose details will be described in the section below.

\subsection{FDM simulation}\label{sec2.1}
In this section, we introduce the detailed setup of the FDM simulation, beginning with the Schr\"odinger-Poisson (SP) equations \cite{Diosi:2014ura,Moroz:1998dh} that govern the behavior of the FDM system.
\begin{align}
i\hbar \frac{\partial}{\partial t} \psi(\mathbf{x},t ) & = -\frac{\hbar^2}{2m}\nabla^2 \psi(\mathbf{x},t )  +m V(\mathbf{x},t ) \psi(\mathbf{x},t ) \label{Schrodinger_Poisson_SP} \\
 \nabla^2 V(\mathbf{x},t )&=4\pi G \, m(|\psi|^2-\langle |\psi|^2 \rangle)(\mathbf{x},t )
\label{Schrodinger_Poisson_P}
\end{align}
The complex wavefunction $\psi(\mathbf{x},t)$ is normalized such that $\int d^3 \mathbf{x}\, m|\psi|^2 = M_{\rm tot} $, indicating that the mass density of the FDM system is $\rho=m|\psi|^2$. 
To simplify the numerical handling of the SP equations, we introduce dimensionless variables by rescaling with a chosen unit-mass scale $\mathcal{M}$, which determines the unit-time $\tau_c$, unit-length $\ell_c$, unit-potential $\alpha_V$, and unit-wavefunction scale $\alpha_\psi$ in our code space.

\begin{align}     
t & \rightarrow   \tau_c 
\, t   = \frac{\hbar^3}{m^3} \frac{1}{\left( G\mathcal{M}\right)^2} t \\
\mathbf{x} &  \rightarrow   \ell_c
\, \mathbf{x}   = \frac{\hbar^2}{m^2} \frac{1}{G\mathcal{M}} \, \mathbf{x}\\
V& \rightarrow   \alpha_V \, V   = \frac{m^2}{\hbar^2} \left({G\mathcal{M}}\right)^{2}  V\\
\psi & \rightarrow   \alpha_\psi \, \psi   =  \left(\frac{\mathcal{M}}{m}\frac{1}{\ell_c^3}\right)^{\frac{1}{2}} \psi
\label{code_unit}
\end{align}
These lead to a dimensionless SP system suited for numerical analysis
\begin{align}   
i \, \frac{\partial}{\partial t}  \psi(\mathbf{x},t ) \, & = -\frac{1}{2}\nabla^2 \psi(\mathbf{x},t )  + V(\mathbf{x},t ) \psi(\mathbf{x},t ), \label{Schrodinger_Poisson_Code_SP}
\\
 \nabla^2 V(\mathbf{x},t )&=\ 4 \pi \, (|\psi|^2-\langle |\psi|^2 \rangle)(\mathbf{x},t ),
\label{Schrodinger_Poisson_Code_P}
\end{align} 
and the normalization of $\psi$ becomes $\int d^3 \mathbf{x}\,  |\psi|^2 = \frac{M_{\rm tot}}{\mathcal{M}} $. 
By taking our unit-mass scale as $\mathcal{M} = 4\pi \times 10^6 M_\odot$ and the FDM particle mass $m=10^{-22}{\ \rm eV}/c^2$, the unit-time and length scales become $\tau_c\simeq 2.358 \ {\rm Gyr}$ and $\ell_c \simeq 6.800\,\rm kpc$.
Accordingly, the unit-velocity scale can be defined as $v_c\equiv \frac{\ell_c}{\tau_c} \simeq 2.819 \rm\,km/s$.

We use the time-independent, spherically symmetric ground-state solution of the SP equations (\ref{Schrodinger_Poisson_SP}) and (\ref{Schrodinger_Poisson_P}) as the initial FDM subhalo configuration, which is commonly referred to as a soliton core \cite{Hui:2016ltb,Schive:2014dra,Hui:2021tkt}.
We represent the size of a soliton with mass $M_s$ by its half-mass radius $r_{1/2} \simeq\,f_0 \frac{\mathcal{M}}{M_s}\ell_c=f_0 \frac{\hbar^2}{GM_s m^2}$ with $f_0\simeq 3.925$ \cite{Hui:2016ltb}.
For $M_s=50\mathcal{M}$, the validity of our superposition approach is evident when considering the size ($\sim 2r_{1/2}\simeq 0.1571\ell_c$) of the solitons compared with the initial distance ($\frac{L}{2}=5\ell_c$) between them.
We denote the wavefunction of this soliton as $h(\mathbf{x}-\mathbf{x}_0;M_s)$ centered at $\mathbf{x}_0$.
A moving solution with an initial velocity $\mathbf{v}_0$ and phase $\varphi_0$ can then be easily obtained \cite{Edwards:2018ccc}:
\begin{equation}
\label{solwave}
\psi_{\rm sol} (\mathbf{x},\mathbf{x}_0; \mathbf{v}_0;M_s;\varphi_0)=h(\mathbf{x}-\mathbf{x}_0;M_s)\, e^{i\mathbf{v}_0 \cdot (\mathbf{x}-\mathbf{x}_0)+i\varphi_0 }.
\end{equation}
Using this, our initial configuration of colliding FDM subhalos, without phase for simplicity, is as follows:
\begin{equation}
\label{twowaves}
\psi (\mathbf{x})= \psi_{\rm sol} (\mathbf{x},\mathbf{x}_L; \mathbf{v}_L;M_s;0) + \psi_{\rm sol} (\mathbf{x},\mathbf{x}_R; \mathbf{v}_R;M_s;0),
\end{equation}

We employ the pseudo-spectral method to solve the SP equations using the open-source python package PyUltraLight \cite{Edwards:2018ccc}.
Since the code assumes periodic boundary conditions, the average density must be subtracted from the right-hand side of the Poisson equation, as shown in (\ref{Schrodinger_Poisson_P}) and (\ref{Schrodinger_Poisson_Code_P}).
Also, the soliton configuration is generated numerically using a fourth-order Runge-Kutta method, as implemented in the file \texttt{soliton\_solution.py} included in the PyUltraLight package, also detailed in \cite{Paredes:2015wga}.
Our computational domain is a box of size $L=10\ell_c$ containing a grid of $N_g=400$ points, designed to optimize initial conditions and computational efficiency with a spatial resolution of $\Delta x=\frac{1}{40}\ell_c\simeq 0.17~\mathrm{pc}$.
The default timestep used in this simulation is chosen as
\begin{equation}
\Delta t_d \equiv \frac{1}{\pi} \left(\frac{\Delta x}{\ell_c}\right)^2 \tau_c
,
\label{default_timestep}
\end{equation}
which becomes $\simeq 1.989 \times 10^{-4} \tau_c$ in our environment.
This ensures numerical stability by preventing the phase difference between two adjacent grid points from exceeding $\pi$, thereby avoiding backward movements in the FDM fluid.
It also limits the maximum velocity in our simulations to $v_{\rm max}=\frac{\Delta x}{\Delta t_d} = \frac{\pi}{\Delta x / \ell_c} v_c \simeq 125.7 v_c$.
Additionally, we found that if $v_i<36v_c$, the two solitons tend to merge, which must be avoided since our aim is to analyze post-collision velocities.
Therefore, we set the range of $v_i$ as $36v_c \leq v_i \leq 124v_c$, resulting in 221 distinct values.
The simulation duration for each setup is $T=\frac{1.4L}{v_i}$, with a total number of snapshots $N_s=2500$. 
The timestep per snapshot $\Delta t=\frac{T}{N_s}$ then becomes $\simeq 1.556\times 10^{-4}\tau_c$ for $v_i=36v_c$, and $\simeq 4.516\times 10^{-5}\tau_c$ for $v_i=124v_c$, both of which are smaller than $\Delta t_d$.
These configurations balance detail and efficiency, making them suitable for detailed collision simulations.

\subsection{CDM simulation}\label{sec2.2} 
In this section, we turn to the CDM setup.
All initial settings, including both the subhalo initial configurations (positions $\mathbf{x}_{L/R}$, velocities $\mathbf{v}_{L/R}$, and mass $M_s$) and the simulation environment (domain size $L$, duration $T$, number of snapshots $N_s$, and periodic boundary conditions), are the same as those described in Section \ref{sec2.1}.
We now prepare an initial CDM subhalo of $M_s=2\pi\times 10^8 M_\odot$ and $r_{1/2}=0.5338\rm kpc$ the same as the FDM subhalo defined earlier, with $N_p=10^5$ identical particles to follow a Hernquist distribution \cite{Hernquist:1990}, defined by
 \begin{equation}
 \rho_{\rm H}(r)=\frac{\rho_0}{\frac{r}{a_s}\left(1+\frac{r}{a_s}\right)^3},
 \label{Hernquist_model}
 \end{equation}
where $a_s$ is the scale radius. 
The gravitational potential $\Phi_{\rm H}(r)$ of this profile, derived from the Poisson equation $\nabla^2 \Phi_{\rm H}=4\pi G \rho_{\rm H}$, becomes
\begin{equation}
\Phi_{\rm H}(r)=-\frac{2\pi G \rho_0 a_s^3}{r+a_s}.
\label{Hernquist_Potential}
\end{equation}
We use the Eddington inversion method \cite{Eddington:1916b, Binney:1993ce} to numerically determine the initial positions and velocities of the particles, deriving the isotropic probability density function from (\ref{Hernquist_model}) and (\ref{Hernquist_Potential}).
The resulting distribution function for (\ref{Hernquist_model}) is given by
\begin{equation}
f_{\rm H}(\mathcal{E}(\mathbf{x}, \mathbf{v}))=\frac{1}{\sqrt{2}(2\pi)^3(GMa_s)^{3/2}} \frac{\sqrt{\mathcal{E}}}{(1-\mathcal{E})^2}\left[(1-2\mathcal{E})(8\mathcal{E}^2-8\mathcal{E}-3)+\frac{3\sin^{-1}\sqrt{\mathcal{E}}}{\sqrt{\mathcal{E}(1-\mathcal{E})}}\right]
\label{Hernquist_Distribution}
\end{equation}
if $\mathcal{E(\mathbf{x}, \mathbf{v})}>0$ and $f_{\rm H}(\mathcal{E(\mathbf{x}, \mathbf{v})})=0$ otherwise, where
$\mathcal{E}(\mathbf{x}, \mathbf{v})=-\frac{a_s}{GM}\left(\frac{1}{2} v^2 - \frac{2\pi G \rho_0 a_s^2}{1+r/a_s} \right)$.
This particle system is initialized within a cut-off radius $R_{\rm cut}=3r_{1/2}$, emulating the compact nature of FDM solitons to facilitate rapid approach to dynamical equilibrium.
The two parameters $\rho_0$ and $a_s$ in (\ref{Hernquist_model}) are determined by the following conditions:
\begin{align}
M_s&=4\pi\int_0^{3r_{1/2}} r^2 \rho_{\rm H}(r)dr \label{Hernquist_Mass1} \\
\frac{1}{2}M_s&=4\pi\int_0^{r_{1/2}} r^2 \rho_{\rm H}(r)dr.
\label{Hernquist_Mass2}
\end{align}
These give $a_s = \frac{3(2\sqrt{2}-1)}{7} r_{1/2}$ and $\rho_0 = \frac{M_s}{\pi r_{1/2}^3}\frac{22+16\sqrt{2}}{27}$.
We perform an $N$-body simulation for $1.5~ \rm Gyr$ to relax the subhalo system into a dynamical equilibrium. 
After relaxation, deviations in both the compactness of the subhalo system and the value of $r_{1/2}$ are negligible.
Therefore, we use the final results as our actual initial subhalo configuration for head-on collision.
Although this system differs from the original Hernquist profile we began with, this does not critically matter in our simulation, because the Hernquist profile does not perfectly estimate the CDM subhalo configuration \cite{Baes:2002tw}.

We use the public code Gadget4 \cite{gadget4} for running a gravitational $N$-body simulation. 
Gadget4 is a parallel cosmological $N$-body code designed for simulations of cosmic and galactic evolution. It computes the nonlinear regime of gravitational dynamics and hydrodynamics. 
In our simulation, we use a TreePM algorithm with the particle mesh grid (PMG) scaled to $\frac{L}{512} \simeq 0.1328\rm kpc$, which optimizes computational efficiency with minimal loss of accuracy compared to the purely Tree algorithm. 
The PMG scaling uses the particle mesh algorithm for larger scales and the tree algorithm for finer details, optimizing both computational speed and precision. 

We set the softening length (SL) scale to the value ${\rm SL} =\epsilon_{\rm acc}\equiv \frac{2r_{1/2}}{\sqrt{N_p}} \simeq 3.377\ \rm pc$ to mitigate the strong discreteness effects \cite{softeninglength}. 
Our choice of ${\rm SL}$ is intended to limit the maximum stochastic acceleration  $a_{\rm max} \sim a_{\rm acc} = Gm_p/ \epsilon_{\rm acc}^2$, which prevents close two-body encounters from producing forces that exceed the mean field acceleration at a radial distance of $2r_{1/2}$ from the center of our subhalo model. 
Particles beyond $R_{\rm cut}$ and any displacement of the COM exert negligible influence on the macroscopic dynamics of our subhalos.
With our choice of SL, a stability test of this particle system is provided in Appendix \ref{A.Test}.
We have also tested various choices around $\epsilon_{\rm acc}$, and concluded that there are no significant differences in the performance of the subhalos.

To facilitate fair comparisons of CDM with FDM, we project the $N$-body particles onto the density map. 
We use the pynbody package \cite{pynbody}, which is a widely used open-source code for analyzing results of astrophysical $N$-body simulations. 
We adjust the resolution to 400 cells per side in the simulation box to balance computational efficiency and accuracy for detecting the core region of subhalos.


\section{Numerical Results}\label{sec3}
\begin{figure}[p]
\begin{center}
\includegraphics[width=0.9\textwidth, trim={3cm 2cm 6cm 2cm}]{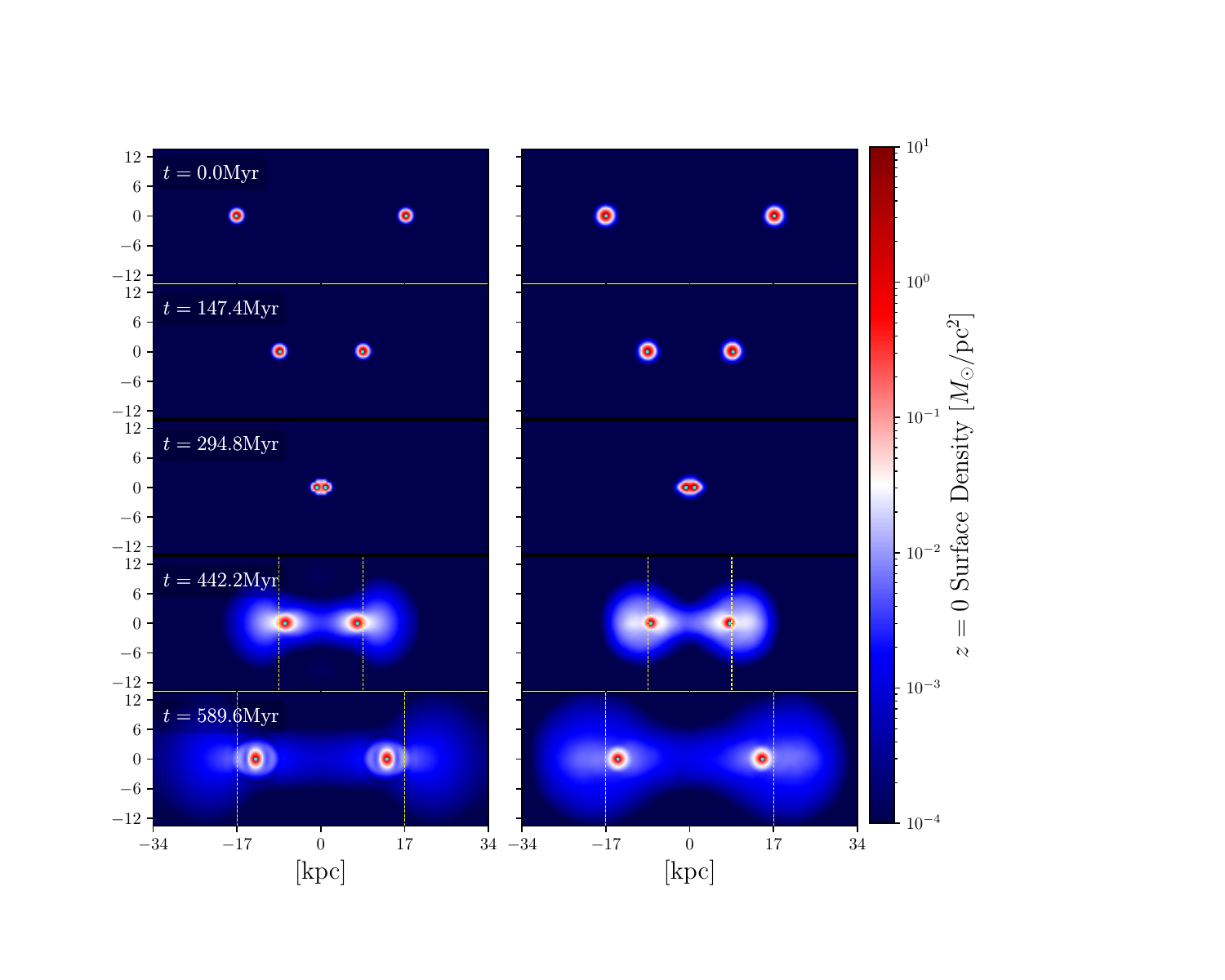}
\caption{
The time evolution of head-on colliding identical subhalos with $v_i=112.8 {\rm km/s}\,(=40v_c)$ for FDM (left) and CDM (right) models.
All density maps of subhalos are sliced along the $z=0$-plane.
The color bar on the right side indicates the surface mass density of unit $M_{\odot}/\rm pc^2$.
The vertical dashed yellow lines in the snapshots of fourth and fifth rows of each panel (after collision), representing $x=\pm 8.5\rm kpc$ and $\pm 17\rm kpc$ respectively, mark the positions of subhalos from the second and first rows.
The cyan dots mark the positions of the subhalos for all snapshots.
The FDM subhalos are observed to be further from snapshots of the dashed lines compared to the CDM subhalos, indicating a greater reduction in post-collisional velocity.}
\label{snap_v40}
\end{center}
\end{figure}

\begin{figure}[p]
\begin{center}
\includegraphics[width=0.9\textwidth, trim={3cm 2cm 6cm 2cm}]{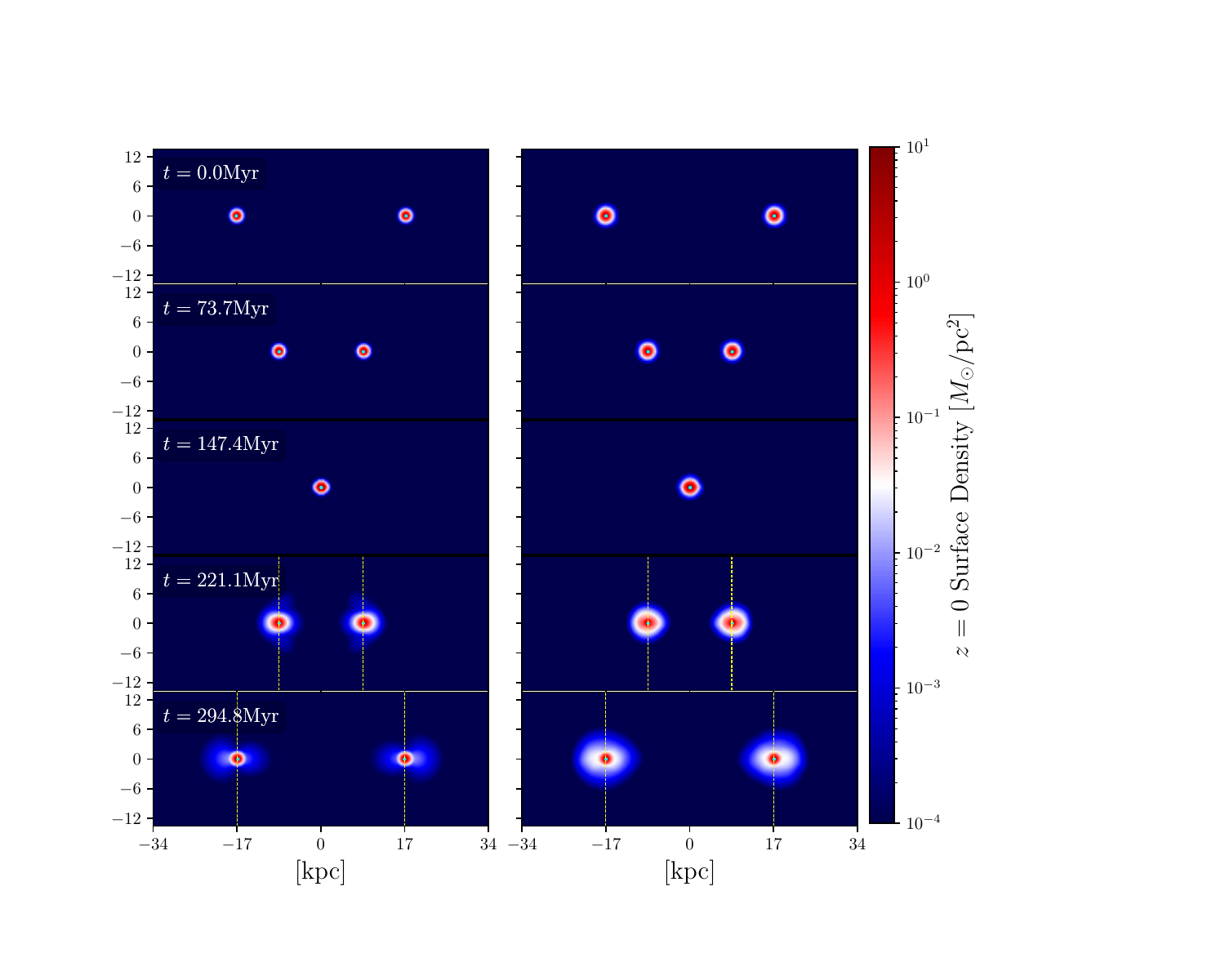}
\caption{
The time evolution of head-on colliding identical subhalos with $v_i=225.6 {\rm km/s}\,(=80v_c)$ for FDM (left) and CDM (right) models.
Snapshots after collision shows subhalos closer to the dashed lines, indicating a smaller reduction in post-collisional velocity compared to Figure \ref{snap_v40}.
}
\label{snap_v80}
\end{center}
\end{figure}

\begin{figure}[htbp]
\begin{center}
\includegraphics[width=0.99\textwidth]{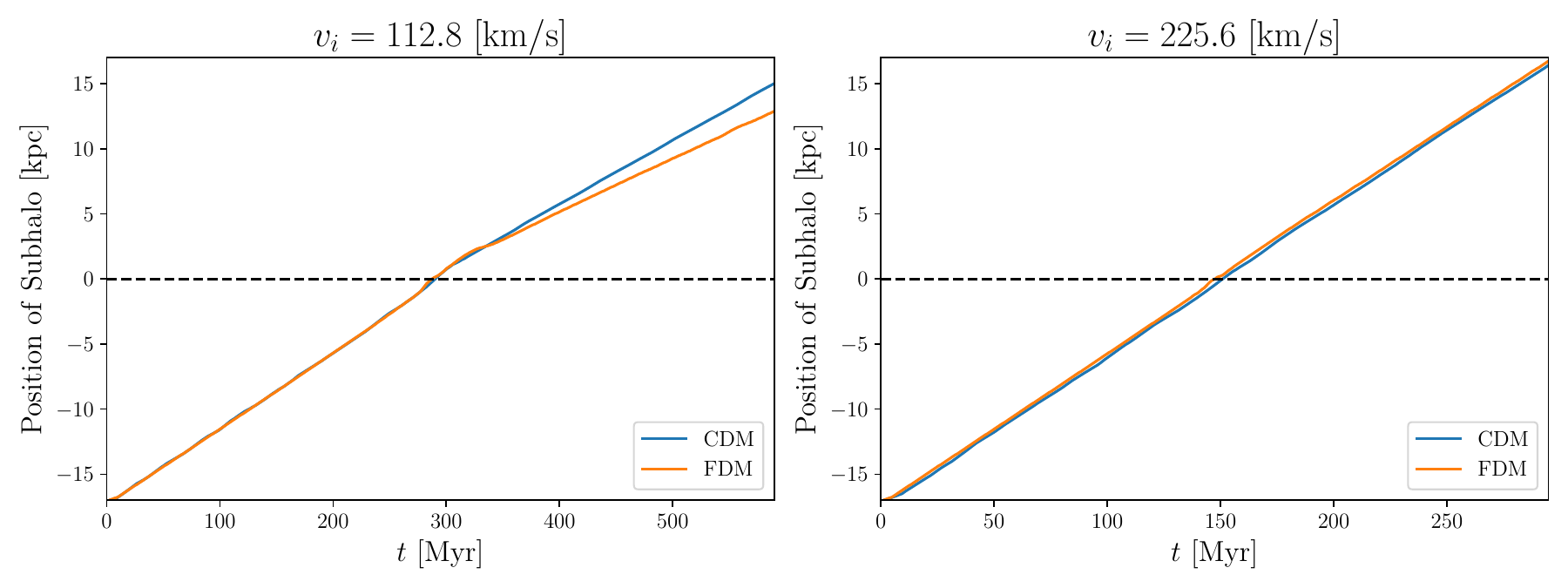}
\caption{
Time evolution of the subhalo position moving along the $+\hat{x}$-direction for both DM models.
The subhalos initially start from $x=-17~\mathrm{kpc}$ with relative velocities $v_i=112.8 \mathrm{km/s}$ (left) and $225.6\mathrm{km/s}$ (right)
The FDM subhalos show a larger velocity reduction after collision compared to the CDM case, especially for the lower velocity scenario.
}
\label{disp_plot}
\end{center}
\end{figure}

This section presents the results of our numerical analysis of head-on collisions between two identical subhalos in both the FDM and CDM models.
Figures \ref{snap_v40} and \ref{snap_v80} display density maps of subhalos in FDM (left) and CDM (right) models, sliced along the $z=0$-plane at various stages of collision.
Figure \ref{snap_v40} displays snapshots taken at five equidistant intervals from 0 to $589.6\rm Myr$, with an initial relative speed of $v_i = 112.8{\rm km/s}\,(=40v_c)$.
Figure \ref{snap_v80} shows similar snapshots for each model over a period from 0 to $294.8\rm Myr$ with $v_i = 225.6{\rm km/s}\,(=80v_c)$.
The selected time intervals align the subhalos at same positions relative to their $v_i$ prior to collision, enabling an effective analysis of post-collisional dynamics.
Furthermore, these figures reveal a marked difference in post-collisional velocity changes, with FDM subhalos exhibiting a significantly greater decrease than their CDM counterparts.
This disparity is more pronounced for lower $v_i$, whereas both models exhibit a smaller velocity decrease for higher $v_i$.
This trend is further supported by Figure~\ref{disp_plot}, which shows the time evolution of the subhalo positions and confirms that the difference between FDM and CDM becomes less significant at higher velocities.

\begin{figure}[htbp]
\begin{center}
\includegraphics[width=0.99\textwidth]{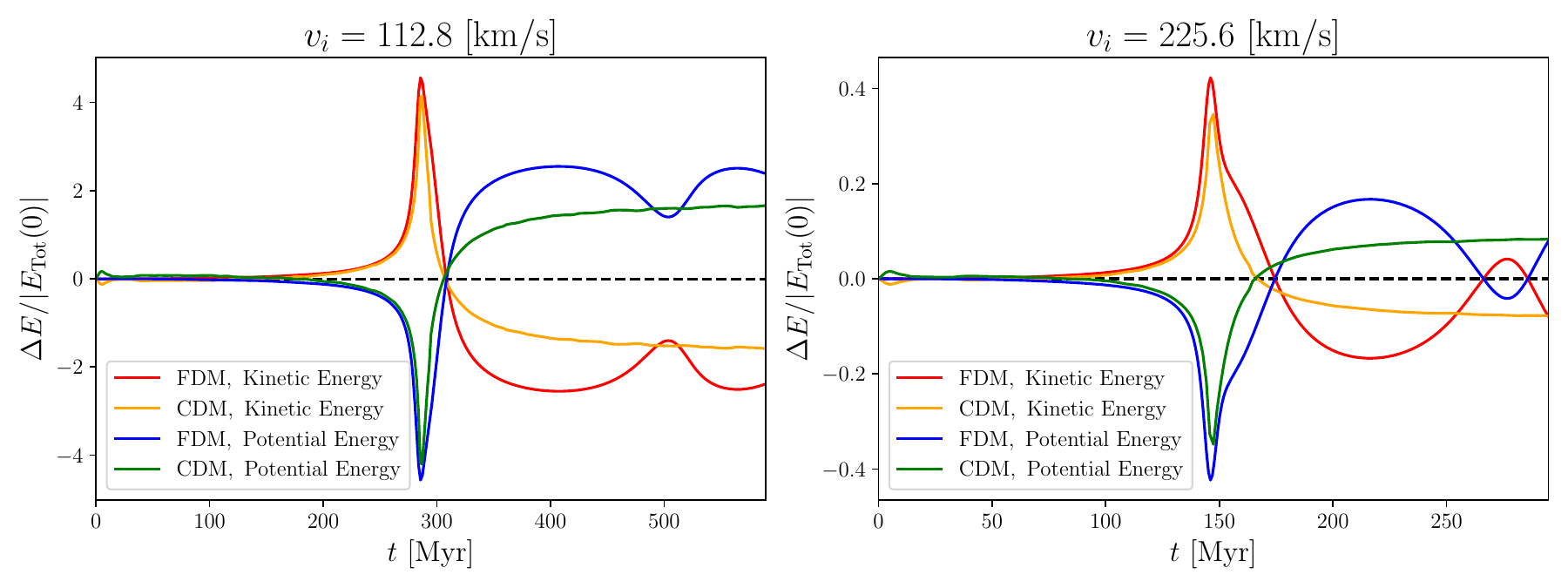}
\caption{
Time evolution of the kinetic/potential energy of the FDM/CDM models, divided by the initial total energy $|E_{\rm tot}(0)|$, for $v_i=112.8 \mathrm{km/s}$ (left) and $225.6\mathrm{km/s}$ (right).
After the collision (at the highest/lowest peak), the kinetic energy of the FDM decreases more significantly than of the CDM system, with some oscillations.
Both of these are manifested by the gravitational cooling effect, particularly at lower $v_i$.
}
\label{energy_plot}
\end{center}
\end{figure}

The individual components of the total energy for both DM models are shown in Figure \ref{energy_plot}. 
First, the FDM system loses kinetic energy more significantly than the CDM system, particularly at lower $v_i$.
Second, both components of the energy of the FDM system exhibit an oscillatory nature, whereas those of the CDM system do not.
Both of these phenomena are based on the gravitational cooling of the FDM system \cite{Guzman:2006yc, Bak:2018pda}.

We now define the location of each subhalo as the position of its density peak.
To calculate the final relative speed $v_f$ of the subhalo, we measure the time it takes for the right-hand subhalo to travel from $x_R' = -\frac{3}{16}L$ to $x_R'' = -\frac{4}{16}L$. The final velocity is then calculated using the formula:
\begin{align}
v_f \equiv 2\frac{x_R'' - x_R'}{t''_R-t'_R},
\end{align}
where $t'_R$ and $t''_R$ are the times at which the subhalo passes through $x_R'$ and $x_R''$, respectively.
This calculation is performed after the subhalo has sufficiently escaped the collision region of the simulation box.
To simplify our analysis, we use $v$ to denote the initial relative speed and $\Delta v$ to denote the change of the relative velocity. 
These relationships are visualized in Figure \ref{v_dv_plot} as dotted data. 
For FDM, $|\Delta v| \sim 51\, {\rm km/s}$ when  $v\sim 100\,{\rm km/s}$ , and $|\Delta v| \sim 2\, {\rm km/s}$ when $v\sim 350\,{\rm km/s}$. 
For CDM, the corresponding values are $|\Delta v| \sim 22\,{\rm km/s}$ and $|\Delta v| \sim 2\,{\rm km/s}$, respectively. 
The FDM (red dotted line) exhibits a larger $|\Delta v|$ than CDM (blue dotted line), particularly at lower $v$.
At higher $v$, the behaviors of FDM and CDM converge significantly.


\section{Analysis}\label{sec4}
In this section, we shall provide a theoretical analysis of $\Delta v$ from the results of DM subhalo collision simulations described in detail in the previous section, separating the FDM and CDM models into two sections.

\subsection{FDM Physics in Head-on Colliding Subhalos}

As discussed in Section \ref{sec1}, both dynamical friction and gravitational cooling influence FDM dynamics.
We begin our analytical studies by following the approach in \cite{effectivecrosssection}.

We denote the distance between the two subhalos as $x_r(t) \equiv x_L(t) - x_R(t)$, setting $x_r(0) = 0$, which means that the collision occurs at $t=0$.
The momentum transfer $\Delta p_L$ to the left soliton from the right, caused by gravitational cooling, is then given by
\begin{equation}
\Delta p_L = \int_{-\infty}^{\infty} dx_r \frac{dt}{dx_r}F(x_r),
\label{Delta_p_L}
\end{equation}
where $F(x_r(t))$ is the effective force on the left soliton along $\Delta p_L$.
In this analysis, we focus on two timescales to explain the perturbation caused by gravitational cooling in our simulations. 
One is the overlapping time interval of the core region of two solitons, regarded as $\Delta \tau_{\rm cross} = \frac{r_{1/2}}{v}$ where $v$ is the relative speed of solitons.
The other is the relaxation timescale in a single FDM soliton of mass $M_s$, given as $\tau=\frac{\hbar^3}{(GM_s)^2m^3}$.
The dissipation fraction during the collision will then be proportional to $\frac{\Delta \tau_{\rm cross}}{\tau}$.
Hence, we modify $\frac{dx_r}{dt}$ as estimated by
\begin{equation}
\frac{dx_r}{dt}\sim \left( \frac{dx_r}{dt}\right)_0 \left[
1-\delta' \frac{\Delta \tau_{\rm cross}}{\tau} \frac{x_r}{r_{1/2}} + O\left(\frac{\Delta \tau^2_{\rm cross}}{\tau^2} \right)
\right],
\label{reduce_velocity}
\end{equation}
with a numerical constant $\delta'\sim \mathcal{O}(1)$, where $\left( \frac{dx_r}{dt}\right)_0$ denotes the velocity before including any dissipative effects such as gravitational interaction.
(We will discuss the validity of the assumption $\Delta \tau_{\rm cross}^2\ll \tau^2$ later.)
By replacing $\frac{dx_r}{dt}$ in (\ref{Delta_p_L}) using (\ref{reduce_velocity}), we get
\begin{equation}
\Delta p_L \propto \int dx_r \left(\frac{dt}{dx_r}\right)_0 
\left[1+\delta' \frac{\Delta \tau_{\rm cross}}{\tau} \frac{x_r}{r_{1/2}} \right]  F(x_r)
\propto -\frac{\Delta \tau_{\rm cross}}{\tau} \frac{GM^2}{r_{1/2}^2}\Delta\tau_{\rm cross}
\label{momentum_transfer_prop}
\end{equation}
with omitting the $\mathcal{O}(1)$ contribution of the integral, which does not affect the evaluation.
Thus, we get
\begin{equation}
\frac{\Delta v}{v}=-\alpha' \left(\frac{\hbar}{mvr_{1/2}}\right)^3
\label{Delta_v_over_v}
\end{equation}
with a numerical constant $\alpha'\sim \mathcal{O}(1)$, using the relation $\frac{\Delta v}{v}=\frac{\Delta p_L}{p_L}$.

Our simulations involve two independent length scales: the half-mass radius of soliton $r_{1/2}$ and the de Broglie wavelength divided by $2\pi$, given by $\lambdabar(v)\equiv \frac{\hbar}{mv}$. 
Using these, we define a dimensionless variable $q$ as follows.
\begin{equation}
q\equiv \frac{\lambdabar(v)}{r_{1/2}}=\frac{\hbar}{mvr_{1/2}}
\label{q_definition}
\end{equation}
In our simulations, where $q$ has a range of $0.1027 \lesssim q \lesssim 0.3538$, we fit our data of $\frac{|\Delta v|}{v}$ using a 2-parameter fit function $G(q)$, which takes the following form:
\begin{equation}
\frac{|\Delta v|}{v}=G(q)=Aq^3(1+Bq^2).
\label{fdm_Gq_fit}
\end{equation}
We find $A=3.692 \pm 0.041$ and $B=15.63 \pm 0.2937$, with the corresponding fit function plotted in Figure \ref{v_dv_plot} as black dashed line.


We briefly discuss the physical meaning of $q$.
First, the region of $q\gg 1 \ (\lambdabar \gg r_{1/2})$ is referred to as the ``classical" regime \cite{Hui:2016ltb, Lancaster:2019mde}.
In this regime, solitons after collision would behave like classical particles when they are well separated.
Second, for the region of $q\ll 1  \ (\lambdabar \ll r_{1/2})$, referred to as the ``quantum" regime \cite{Hui:2016ltb, Lancaster:2019mde}, the wave nature of the solitons is more pronounced.
In this regime, solitons smoothly pass through each other like typical colliding wave packets.
The values of $q$ in our simulations lie in the transition region between the classical and quantum regimes.
These also satisfy $q^2 <1$, indicating that the assumption $\Delta \tau_{\rm cross}^2\ll \tau^2$ for (\ref{reduce_velocity}) is valid in our simulations.
From the fit function $G(q)$ in (\ref{fdm_Gq_fit}) and the fitting parameters, we find that when $q>q_{\rm mer}\simeq 0.3568$, corresponding to $v_{\rm mer}\simeq 100.7\rm km/s$, the function $G(q)= \frac{|\Delta v|}{v}$ exceeds $1$, indicating that solitons merge rather than scatter.
Thus, the classical regime in our context only describes the merging process of solitons.

Dynamical friction \cite{Lancaster:2019mde} is also expected to be a relevant mechanism in our simulations.
First, we consider an object of mass $M_{\rm src}$ with a radius scale $\ell_{\rm size}$, traveling through a DM field of density $\rho$ at a speed $v$.
For a point-mass object ($\ell_{\rm size}=0$), the dynamics can be viewed as Coulomb scattering, with the details analytically calculated in \cite{Hui:2016ltb, Lancaster:2019mde}, and numerically tested in \cite{Wang:2021udl}.
When the object travels through the DM field, a wake is formed behind it is due to gravitational interactions.
This effective dragging force acts as dynamical friction, given by
\begin{equation}
F = -4\pi\rho \left(\frac{GM_{\rm src}}{v}\right)^2 C \left(q_{\rm src}, q_r, q_{\rm size}\right)
\label{fdm_df}
\end{equation}
where $C(q_{\rm src},q_r,q_{\rm size})$ is a dimensionless coefficient that is a function of three dimensionless variables--$q_{\rm src}=\frac{\lambdabar(v)}{r_{1/2,\rm src}}$ with $r_{1/2,\rm src}=f_0 \frac{\hbar^2}{GM_{\rm src}m^2}$, $q_r = \frac{\lambdabar(v)}{r}$, and $q_{\rm size} = \frac{\lambdabar(v)}{\ell_{\rm size}}$. 
Applying this to our context, we replace the object with a soliton of mass $M_{\rm src}$, the DM field with another soliton of mass $M$, and their relative speed $v$.
Thus, we naturally take $q_r \sim q_{\rm size} \sim q$.
Since the frictional force $F$ acts roughly for $\Delta \tau_{\rm cross}= \frac{r_{1/2}}{v}$, its contribution to $\frac{|\Delta v|}{v}$ may be estimated as
\begin{equation}
\frac{|\Delta v|}{v}\bigg|_{\rm df,FDM} \sim \frac{3M_{\rm src}}{2M}f_0^2 q^4C(q_{\rm src},q,q)
\label{fdm_dv_df}
\end{equation}
where we take the average density of the soliton as $\rho\sim \frac{1}{2}M/\left(\frac{4}{3}\pi r_{1/2}^3\right)$, which respects the definition of the half-mass radius $r_{1/2}=f_0 \frac{\hbar^2}{GMm^2}$.
For the point-mass object ($q_{\rm size} \rightarrow \infty$) and for $q_{\rm src}\ll 1$, the expression of $C$ becomes \cite{Hui:2016ltb, Lancaster:2019mde,Wang:2021udl}
\begin{equation}
C(q_{\rm src}, q_r, \infty)={\rm Cin}\left(\frac{2}{q_r}\right)+\frac{\sin (2/q_r)}{2/q_r}-1+\mathcal{O}(q_{\rm src}^1)
\label{Coefficient_for_Classical_Limit}
\end{equation}
where ${\rm Cin}(x)=\int_0^xdt\frac{1-{\rm cos}\,t}{t}$.
For our values of $q$ in the range of $0.1027 \lesssim q \lesssim 0.3538$, $C(q,q,\infty)$ may be estimated as $1.326\lesssim C(q,q,\infty)\lesssim 2.548$. Thus, (\ref{fdm_dv_df}) makes a significant contribution to the change of the relative velocity, rather than the $q^5$ term in (\ref{fdm_Gq_fit}).
However, the finite size of our object soliton ($\ell_{\rm size}\sim r_{1/2}$) reduces the effect by at least a factor of $10^{-2}$ compared to $C(q,q,\infty)$ in our results, which is treated as a Plummer-like object in Figure 3 of \cite{Lancaster:2019mde}.

We fit our simulation results with several fitting functions to numerically check our above claim.
First, the 2-parameter fit function $G_{1}(q) = Aq^3 (1+B_{\rm df}q)$ fails to optimize.
Second, we attempt the 3-parameter fit with $G_{2}(q) = Aq^3(1+B_{\rm df}q+Bq^2)$, and obtain $B_\mathrm{df} -2.338 \pm 0.0622$. 
Third, noting that $\mathrm{Cin}(x)+\frac{\sin x}{x}\simeq \log x+\gamma_{\rm E}$ for large $x$ ($\gamma_{\rm E} \simeq 0.5772$ is the Euler-Mascheroni constant) \cite{Hui:2016ltb}, we also attempt the 3-parameter fit with $G_{3}(q) = Aq^3(1+B'_{\rm df}q \log q^{-1} +Bq^2)$, and find $B'_{\rm df}=-1.274 \pm 0.0195$.
The negative coefficients found in $G_2(q)$ and $G_3(q)$ are not allowed according to the definition for dynamic friction.
These results suggest that the contribution of dynamical friction to the change of the relative velocity is negligible.


\subsection{CDM Physics in Head-on Colliding Subhalos}
In this section, we focus on the CDM subhalo collision, which is governed solely by classical Newtonian dynamics.
The basic mechanism of dynamical friction for particle systems has already been introduced in the previous section.
However, unlike FDM, the half-mass radius $r_{1/2}$ of a CDM subhalo with mass $M$ is determined by two parameters, $\rho_0$ and $a_s$, as derived from equations (\ref{Hernquist_Mass1}) and (\ref{Hernquist_Mass2}).
Additionally, we introduce two additional length scales--$\ell_\sigma \equiv \frac{GM}{\sigma^2}$ and $\ell_v \equiv \frac{GM}{v^2}$.
Here, $\sigma$ is the typical velocity dispersion of the isotropic subhalo, chosen as $\sigma=\sqrt{\frac{GM}{R_{\rm cut}}}\simeq 50.31 \mathrm{km/s}$ where $R_{\rm cut}=3r_{1/2}$.
First, $\ell_{\sigma}$ is considered a typical length scale of the subhalo constituents.
Second, $\ell_v$ corresponds to an impact parameter at which the deflection angle becomes $90^{\circ}$ when the collision velocity of subhalos $v$ is much larger than $\sigma$.
We expect that this affects the change of the relative velocity of CDM subhalos after the collision.

\begin{figure}[htbp]
\begin{center}
\includegraphics[width=0.65\textwidth]{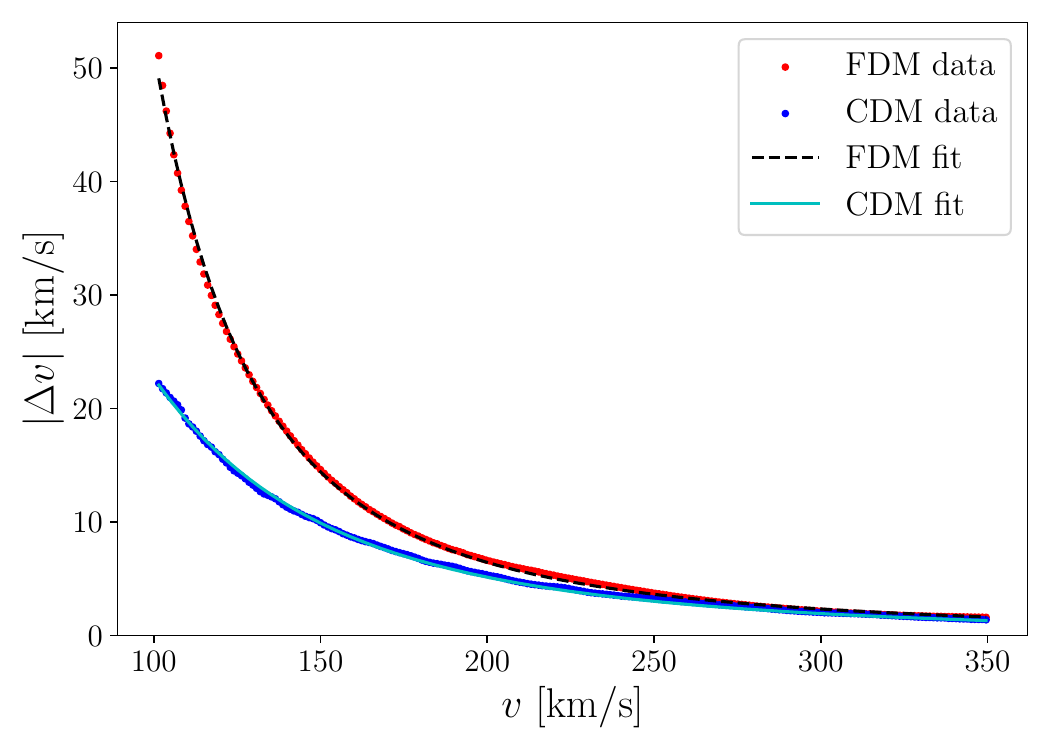}
\caption{Change of the relative velocity after collision $\Delta v$ for FDM (red dots) and CDM (blue dots) models, plotted as a function of the initial relative speed $v$ of 221 data points.
For higher $v$, the post-collisional behavior of both DM models are almost consistent.
However, for lower $v$, $\Delta v$ of FDM model is larger than of CDM model.
This indirectly exhibits the dynamical difference of two DM models, particularly at lower $v$.
Also, we perform 2-parameter fits for the FDM and CDM models using the fitting functions $\frac{|\Delta v|}{v}=Aq^3(1+Bq^2)$ with dimensionless variable $q=\frac{\hbar}{mvr_{1/2}}$ (black dashed line), and $\frac{|\Delta v|}{v}=\bar{A}\bar{q}^2(-\log \bar{q}+B)$ with $\bar{q}=\frac{GM}{v^2 r_{1/2}}$, respectively (cyan solid line)
The estimated dimensionless parameters are $(A,~B)=(3.692 \pm 0.0418,~15.63 \pm 0.2937)$ for FDM and $(\bar{A},~\bar{B})=(0.5168 \pm 0.0041,~1.030 \pm 0.0161)$ for CDM, respectively.
}
\label{v_dv_plot}
\end{center}
\end{figure}

We express the dynamical friction force of our results as
\begin{equation}
F=-4\pi \rho \left(\frac{GM}{v}\right)^2 \bar{C}(\Lambda, \Lambda_{v} , \Lambda_{\rm size}),
\label{cdm_df}
\end{equation}
where $\bar{C}(\Lambda, \Lambda_{v} , \Lambda_{\rm size})$ is a dimensionless coefficient that is a function of three dimensionless variables--$\Lambda=\frac{r_{1/2}}{\ell_{\sigma}}$, $\Lambda_{v}=\frac{\ell_v}{\ell_{\sigma}}$, and $\Lambda_{\rm size}=\frac{\ell_{\rm size}}{\ell_{\sigma}} \simeq 3\Lambda $. 
Applying this to our context, similar to the previous section with $\Lambda_{\rm size}  \sim \Lambda$ and the overlapping time interval defined similarly as before: $\Delta \tau_{\rm cross} = \frac{r_{1/2}}{v}$, the contribution of dynamical friction to $\frac{|\Delta v|}{v}$ may be estimated as
\begin{equation}
\frac{|\Delta v|}{v}\bigg|_{\rm df,CDM} \sim \frac{3}{2}\left(\frac{\Lambda_v}{\Lambda}\right)^2 \bar{C}(\Lambda, \Lambda_v, \Lambda)
\label{cdm_dv_df}
\end{equation}
where we take $\rho\sim \frac{1}{2}M/\left(\frac{4}{3}\pi r_{1/2}^3\right)$.
In general, for Maxwell-like velocity distribution of the DM field, the expression of $\bar{C}$ becomes \cite{Chandrasekhar:1943ys, Binney:1993ce}
\begin{equation}
\bar{C}(\Lambda, \Lambda_{v} , \Lambda_{\rm size}) \sim \log \Lambda_{\rm size} \left[\mathrm{erf}(X) - \frac{2X}{\sqrt{\pi}}e^{-X^2}\right],
\label{Coefficient_for_CDM}
\end{equation}
where $X=\frac{v}{\sqrt{2}\sigma}=(2\Lambda_v)^{-1/2}$ with $\mathrm{erf}(x)=\frac{2}{\sqrt{\pi}}\int_{0}^{x}dt e^{-t^2}$, and $\log \Lambda_{\rm size}$ is the Coulomb logarithm.
In our simulation data, terms in square brackets remain almost 1.
To estimate the contribution of the deflection effect due to the collision on the Coulomb logarithm, we first define a dimensionless variable $\bar{q}$ as follows:
\begin{equation}
\bar{q}\equiv \frac{\Lambda_v}{\Lambda}=\frac{GM}{v^2 r_{1/2}}.
\label{qbar_definition}
\end{equation}
In our simulations, where $\bar{q}$ ranges from $0.0414\lesssim \bar{q} \lesssim 0.4915$, we fit our data of $\frac{|\Delta v|}{v}$ using a 2-parameter fit function $H(\bar{q})$, which takes the following form:
\begin{equation}
\frac{|\Delta v|}{v}=H(\bar{q})=\bar{A} \bar{q}^2 \left(  \log \frac{1}{\bar{q}} + \bar{B} \right).
\label{cdm_HL_fit}
\end{equation}
We find $\bar{A}=0.5168 \pm 0.0041$ and $\bar{B}=1.030 \pm 0.0161$, with the corresponding fit function plotted in Figure \ref{v_dv_plot} as cyan solid line.

We briefly discuss the physical meaning.
First, during the collision, each subhalo's relative speed $v$ acts as the velocity dispersion due to $v\gg \sigma$.
This amplifies the dissipation of the subhalo constituents, leading us to replace the Coulomb logarithm $\log \Lambda_{\rm size}\equiv \log \frac{\ell_{\rm size}}{GM/\sigma^2}$ of effective dimensionless coefficient with $\sim \log \frac{\ell_{\rm size}}{GM/v^2}$.
Nevertheless, this slight increase in the term does not significantly affect $|\Delta v|$, because the dominant term in (\ref{cdm_HL_fit}) is obviously $\sim \bar{q}^2=\left(\frac{GM}{v^2 r_{1/2}}\right)^2$.
Second, the finite size of our Hernquist-like object reduces the Coulomb logarithm by approximately 1.5 compared to point-like results \cite{Esquivel:2007sv}.
These results are analogous to those from FDM.

\section{Conclusions}\label{sec5}

In our study, we have numerically shown that the dissipation after a head-on collision of two DM subhalos is greater in FDM models compared to CDM.
Using distinct numerical methods for each DM model, our simulations revealed that FDM subhalos experience a greater change in relative velocity after collision than CDM subhalos, particularly at lower initial velocities.
Our analysis confirms that the unique behavior of subhalos in FDM, characterized by significant velocity changes, can be accurately modeled by gravitational cooling---a mechanism that stabilizes by dissipating kinetic energy, distinct from the classical dynamical friction observed in CDM.
These findings underscore the differences in subhalo collisions between two DM models, offering deeper insights into how FDM can resolve longstanding discrepancies in DM observations.
Notably, the Bullet cluster, with its relatively high collisional velocity of approximately $4700 \rm km/s$, is consistent with CDM predictions \cite{bullet}.
In contrast, Abell 520, with its relatively low collisional velocity of approximately $1000 \rm km/s$, provides evidence against CDM predictions \cite{abell520}.
Our numerical results align with these observational data, highlighting substantial implications for the broader understanding of DM interactions.

However, our simulations are conducted under idealized assumptions.
Future research could explore a wider range of conditions, such as varying impact angles, mass ratios, and the inclusion of self-interaction among FDM particles \cite{Glennon:2020dxs}.
Higher-resolution simulations, such as those employing adaptive mesh refinement (AMR) grid \cite{Schwabe:2020eac}, could also offer more detailed insights into the microphysical processes involved in DM interactions.
Nevertheless, the results of our work offer a foundation for the continued investigation of FDM, particularly in contexts that challenge traditional CDM models.

In conclusion, our study highlights the potential of FDM not just as an alternative to CDM in theoretical models but as a practical solution to astrophysical puzzles, such as the colliding galaxy clusters.
The distinct energy dissipation mechanisms in FDM, compared to CDM, result in markedly different post-collisional dynamics, providing a robust framework that could explain the complex and seemingly contradictory behaviors of DM observed in galaxy cluster collisions.
These findings suggest that FDM could play a crucial role in resolving longstanding discrepancies in DM observations, offering new avenues for understanding the true nature of DM in the universe.
Continued exploration, supported by both theoretical and observational efforts, will be essential in further validating the FDM model and its implications for DM research.

\subsection*{Acknowledgement}\label{acknowledgement}
We would like to thank Prof. Jae-Weon Lee for enlightening comments, and Prof. Dongsu Bak and Dr. Sangnam Park for the collaboration at the initial stage of this work.

\subsection*
{Funding}\label{funding}
This work was supported in part by Basic Science Research Program through National Research Foundation (NRF) funded by the Ministry of Education
(2018R1A6A1A06024977).

\begin{appendix}
\section{Some tests for the CDM subhalo particle system} \label{A.Test}
This appendix briefly presents the results of a standalone simulation performed to verify the stability of the CDM subhalo described in Section \ref{sec2.2}, prior to initiating the head-on collision analysis.

\begin{figure}[htbp]
\begin{center}
\includegraphics[width=0.99\textwidth]{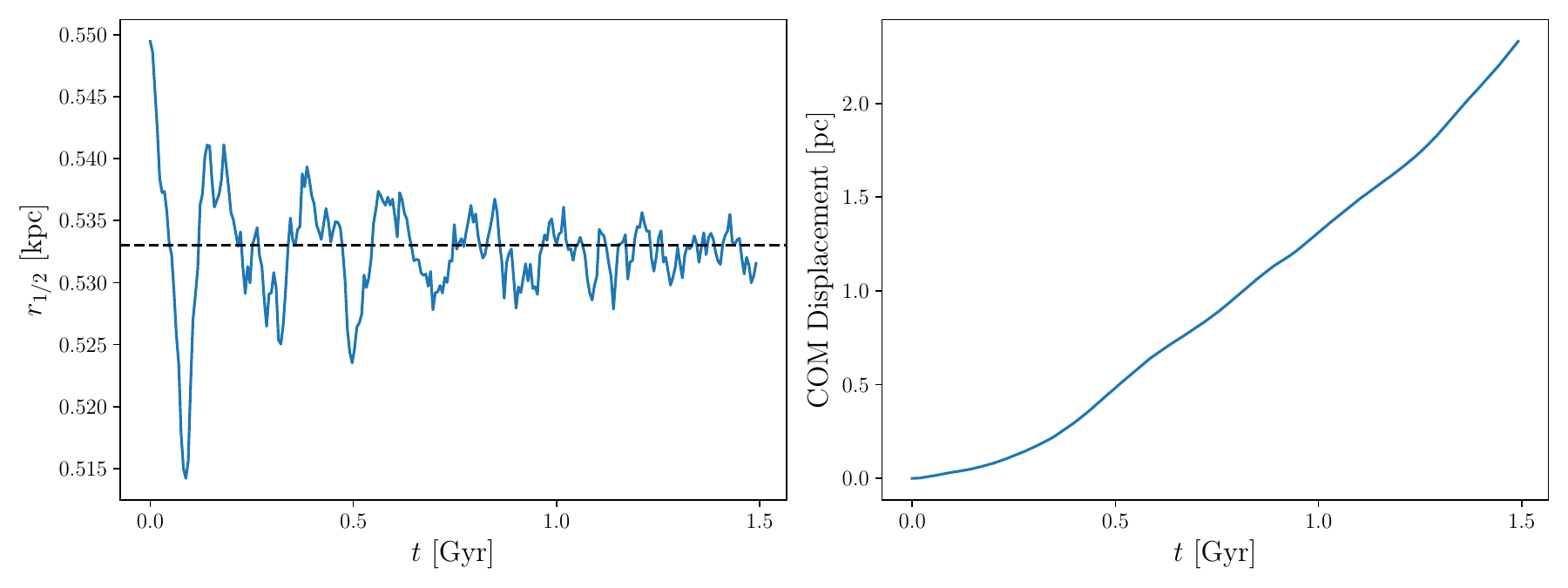}
\caption{
Time evolution of the initial particle system over $1.5~\mathrm{Gyr}$.
(Left) Half-mass radius $r_{1/2}$ starting from $0.55~\mathrm{kpc}$ stabilizes around $0.533~\mathrm{kpc}$ (black dashed).
(Right) COM drifts by approximately $2.34~\mathrm{pc}$ from the origin.
}
\label{check}
\end{center}
\end{figure}

Our initialized particle system consists of $N=10^5$ particles with a total mass $M=2\pi \times 10^8 M_\odot$, a half-mass radius $r_{1/2}=0.55~ \mathrm{kpc}$, a cut-off radius $R_\mathrm{cut}=3r_{1/2}$ and a softening length $\mathrm{SL}=3.377~ \mathrm{pc}$.
First, during the $1.5~\mathrm{Gyr}$ time evolution, the half-mass radius $r_{1/2}$ stabilizes around $0.533~\mathrm{kpc}$, exhibiting only small oscillations.
This value closely matches that of an FDM soliton with the same mass.
Second, the COM drifts by approximately $2.34~\mathrm{pc}$ from the origin, which is negligible compared to the resolution scales of our simulations—namely, the TreePM algorithm’s PM grid size of $132.8~\mathrm{pc}$ in the CDM case and the unit grid size of $170~\mathrm{pc}$ in the FDM case.
The time evolutions of both $r_{1/2}$ and the COM displacement are plotted in Figure~\ref{check}.

\begin{figure}[htbp]
\begin{center}
\includegraphics[width=0.65\textwidth]{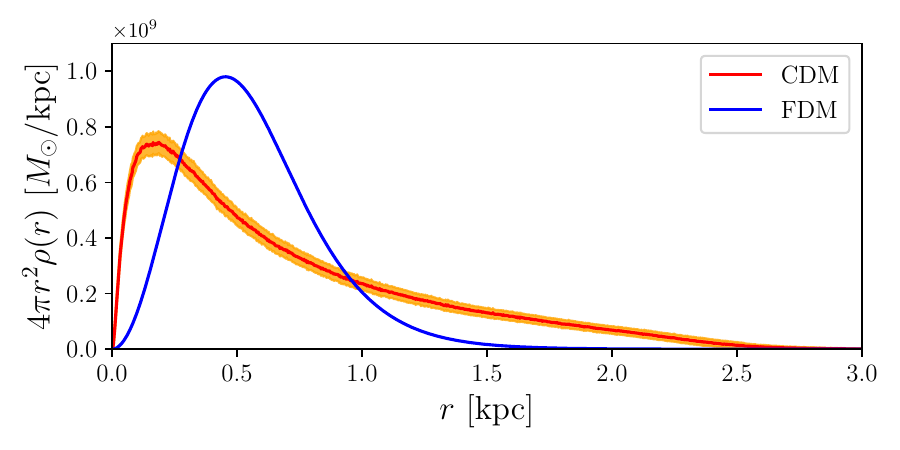}
\caption{
Radial mass distribution $4\pi r^2 \rho(r)$ of the CDM subhalo (red: time-averaged over 300 snapshots; orange region: 1$\sigma$ standard deviation), compared with that of an FDM soliton of the same mass (blue).
The narrow spread confirms the stability of the CDM profile, and the similarity in compactness demonstrates that it is well-matched to the FDM counterpart.
}
\label{check2}
\end{center}
\end{figure}

To further assess the stability of the CDM subhalo, we examined the time evolution of the radial ($r$) distribution of particles from the COM.
Specifically, we computed histograms of particle counts in 1000 radial bins within $3 ~\mathrm{kpc}$--corresponding to a bin width of $3~\mathrm{pc}$--for all 300 snapshots taken over the full $1.5~\mathrm{Gyr}$ simulation.
These histogram curves are proportional to $4\pi r^2 \rho(r)$, where $\rho(r)$ is the mass density.
Figure~\ref{check2} shows the radial distribution of CDM subhalo, with the time-averaged profile and 1$\sigma$ standard deviation over time, along with the corresponding FDM soliton profile computed using \texttt{soliton\_solution.py} included in the PyUltraLight package.
The consistency of the CDM distribution over time demonstrates the robustness and equilibrium of our initialization.
Moreover, the close agreement in compactness between the CDM and FDM profiles confirms that the CDM subhalo is sufficiently compact and well-prepared for comparison.

\end{appendix}

\end{document}